\documentclass[aps,showpacs,preprintnumbers,amsmath,amssymb, 
twocolumn, tightenlines,
]{revtex4}

\usepackage[dvips]{graphicx}
\usepackage{bm}

\begin{document}

\bibliographystyle{apsrev} 

\title {Radiative corrections in fermion bags bound by Higgs boson exchange}

\author{M. Yu. Kuchiev} \email[Email:]{kmy@phys.unsw.edu.au}
\author{V. V. Flambaum} \email[Email:]{flambaum@phys.unsw.edu.au}

\affiliation{School of Physics, University of New South Wales, Sydney
  2052, Australia}

%

    \date{\today}

    \begin{abstract} 
Radiative corrections for several heavy fermions bound together via the Higgs boson exchange are studied. The fermion bags considered include 12, or fewer, fermions occupying the lowest $S_{1/2}$ shell. It is shown that for `moderately heavy' fermions with masses $0.4\lesssim m c^2\lesssim 1$ TeV the radiative corrections are small, $\sim 10^{-2}$, 
and have an attractive nature. 
Therefore they do not put existence of the fermion bag in doubt. This proves that these fermion bags can exist in nature.
    \end{abstract}


%

    \maketitle


Higgs boson exchange induces attractive interaction between fermions, which strength increases with the fermion mass. One can expect therefore that several heavy fermions can form a bound state, a bag of fermions
 \cite{%
Vinciarelli:1972zp,%
PhysRevD.9.2291,%
Chodos:1974je,%
Creutz:1974bw,%
Bardeen:1974wr,%
Giles:1975gy,%
Huang:1975ih,%
PhysRevD.15.1694,%
PhysRevD.25.1951,%
PhysRevLett.53.2203,%
PhysRevD.32.1816,%
Khlebnikov:1986ky,%
Anderson:1990kb,%
MacKenzie:1991xg,%
Macpherson:1993rf,Johnson:1986xz}. 
Refs.\cite{Froggatt:etal} suggested looking at the magic number, $12$ fermions $=6$ tops $+\,6$ antitops occupying the $S_{1/2}$ shell with 3 colors. The estimation there seemed to indicate that such a system should form a tightly-bound state, but calculations of \cite{Kuchiev:2008fd} supported by \cite{Richard:2008uq} showed that this system is unbound. 
For an opportunity to create the bag with many particles see \cite{Crichigno,Kuchiev:2008gt,Crichigno:2010ky}.
In \cite{Kuchiev:2008gt} it was shown 
that for fermions substantially heavier than the top quark the bag made from several fermions can be created. The lower boundary for the fermion mass, which makes formation of a bag possible, 
was found in \cite{Kuchiev:2010ux} to be $320-410$ GeV for the 12 fermion system, where the spread reflects an uncertainty in the Higgs mass $m_\text{H}=100 - 200$ GeV. 
 Thus, a feasible opportunity to observe a bag constructed from several fermions relies on the expectation that there exist heavy fermions of the next, forth generation. Such heavy  fermions with masses 300-500 GeV have  been suggested recently to explain new results on CP violation in B decays \cite{B}. Note also
that the bags with heavy fermions may play a crucial role in the bariogenesis
\cite{Kuchiev:2008gt,Flambaum:2010}.

The challenging issue is that 
 for large fermion masses the radiative corrections resulting from the fermion-Higgs interaction seem also to be enhanced. Previous estimations indicated that these corrections may destabilize the vacuum
 \cite{Dimopoulos:1990at,Bagger:1991pg,Farhi:1998vx}. This may happen for the
 fermion mass $ m > 2$ TeV. 
 We show that for the fermion masses  smaller than 1 TeV
these corrections remain definitely small
and increase the attraction.
Therefore, the fermion bag can exist in nature.

 To calculate the radiative corrections we use the method which was developed
 to calculate the Lamb shift in atoms - see e.g. \cite{QED}. It turned out that technically the problem for the Higgs bag is simpler since the Higgs finite mass
cuts off the infrared divergence.
 We express the radiative correction in terms
 of the renormalized Higgs vertex correction
 (which gives the fermion self-energy)
and the renormalized Higgs polarization operator (the fermion loops).
To explain and support our numerical results we present transparent
analytical estimates, which are valid for several fermions in the bag, $2\le N\le 12$, and provide 
qualitative
comparison with the results of the atomic Lamb shift calculations.   

Consider fermions interacting with the Higgs doublet $\Phi$. Take  the conventional unitary gauge when $\Phi$ is represented by the real field $ \xi$,
$\Phi=\frac{v} { \sqrt {2} }\,( 0 , \xi )^T$, where $v=246$ GeV is the Higgs VEV. The part of the Standard Model Lagrangian, which describes interacting Higgs and fermion fields reads 
\begin{align}
{\cal L}=
\frac{v^2}{2}\Big(\partial^\mu {\xi}\, \partial_\mu {\xi}-\frac{m_{\text H}^2}{4}(\xi^2-1)^2\Big)\!+
\bar \psi\,( \,\hat p-gv \xi \,)\, \psi\,,
\label{L}
\end{align}
where $g$ is defined by the fermion mass $g=m/v$. Following \cite{Kuchiev:2008gt,Kuchiev:2010ux} we use the relativistic mean field approximation as an initial step in the construction of a bag of $N$ fermions occupying the same shell with orbital momenta $j,l$ having in mind the case of $N=12$ fermions in $S_{1/2}$ shell.
Presume that $\xi(\boldsymbol{r})$ and $\psi(\boldsymbol{r})$ are the wave functions, which represent the Higgs and fermions in the mean field approximation. The Hamiltonian, which follows from (\ref{L}), 
can be presented as a sum $H=H_\xi+H_\psi$, where $H_\xi$ describes the pure Higgs field, while $H_\psi$ accounts for fermions interacting with the Higgs
\begin{align}
&H_\xi\,=\,\frac{v^2}{2}\,\int
\Big(\, (\boldsymbol{\nabla}\xi)^2+\frac{1}{4}\,m_\text{H}^2\,(\xi^2-1)^2
\,\Big)\,d^3r~,
\label{Hxi}
\\
&H_\psi \,=\, N\, \langle\,\psi^\dag|\,-i\boldsymbol{ \alpha }\cdot \boldsymbol{\nabla}+ \,g\,v\,\xi\,\gamma_0\,|\,\psi \,\rangle~.
\label{HFG}
\end{align}
The resulting self-consistent equations of the mean field approximation for a spherically symmetrical bag read
\begin{align}
&\!\!\Delta \xi(r)+\frac{m_\text{H}^2}{2}\xi(r)(1-\xi^2(r))=
\frac{(N-1)m}{v^2}~\bar \psi(\boldsymbol{r})\psi(\boldsymbol{r}),
\label{xi}
\\
&\varepsilon\, \psi(\boldsymbol{r})\,=\,
(-i \boldsymbol{ \alpha } \cdot \boldsymbol{\nabla}+ \,g\,v\,\xi(r)\,\gamma_0\, )\,\psi(\boldsymbol{r})~.
\label{F}
\end{align}
Here $\psi(\boldsymbol{r})$ can be conventionally presented via 
its large $f(r)=F(r)/r$ and small $g(r)=G(r)/r$ radial components. 
\begin{figure}[b]
\centering
\includegraphics[height=4.5 cm,keepaspectratio = true, 
]{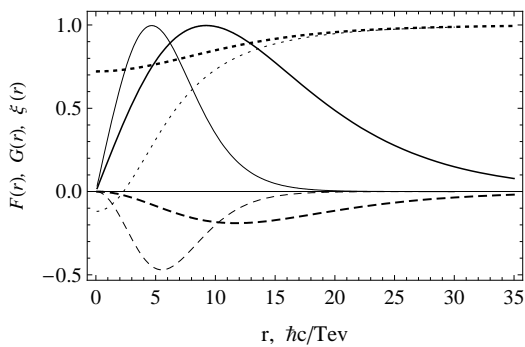}
\caption{
 \label{odin} The wave functions of the fermion bag 
for $N=12$, $j=1/2$, and $m_\text{H}=0.1$ TeV;
solid, dashed and dotted lines - $F(r),G(r)$ and $\xi(r)$ from Eqs. (\ref{xi}), (\ref{F}); three thick lines - the fermion mass $m=350$ GeV,
three thin lines - $m=600$ GeV; $F(r)$ and  $G(r)$ 
are scaled to satisfy $\max\big( F(r)\big)=1$.
 }\end{figure}
\noindent
Fig. \ref{odin} shows behaviour of solutions of Eqs.(\ref{xi}), (\ref{F}), which describe
the bag of $N=12$ fermions in $S_{1/2}$ shell for  $m_\text{ H}=100$ GeV with the fermion mass being chosen either $m=350$ GeV, or $m=600$ GeV.  Similar solutions are found for any fermion mass above the threshold value, $m\ge m_\text{th}$, which in the case at hand is $m_\text{th}\approx 320$ GeV \cite{{Kuchiev:2010ux}}. 

\begin{figure}[t]
\centering
\includegraphics[height=4.5 cm,keepaspectratio = true, 
]{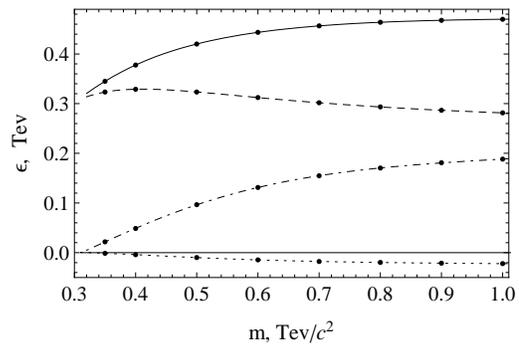}
\caption{
 \label{dva} Energy per fermion $\epsilon=E/N$ vs fermion mass $m$, the fermion bag is same as in Fig. \ref{odin}: solid, dashed, dot-dashed, dotted lines  - total energy $E$ from $H=H_\psi+H_\xi$, fermion contribution $H_\psi$ (\ref{Hxi}), Higgs contribution $H_\xi$ (\ref{HFG}), non-linear term 
 $H_{\xi, \text{\,n-l}}$  (\ref{Hnonlin}); bold dots - calculations, lines - interpolation.}
 \end{figure}

Fig. \ref{dva} presents the energy of the bag per fermion $\epsilon=E/N$, where the total energy $E$ is found from Eqs. (\ref{Hxi}), (\ref{HFG}), in which $\xi$ and $\psi$ satisfy (\ref{xi}), (\ref{F}).  The energy of the fermion term (\ref{Hxi}) greatly exceeds the energy of the Higgs (\ref{HFG}) for small fermion masses, while with the mass increase they become closer. If we split the Higgs term into its linear and nonlinear parts, $H_\xi=H_{\xi,\text{\,lin}}+H_{\xi,\text{\,n-l}}$, 
\begin{align}
&H_{\xi,\text{\,lin}} =\frac{1}{2}\,v^2\int \big( \,(\boldsymbol{\nabla}\xi)^2 +m_\text{H}^2 (\xi-1)^2\big)\,d^3r~,
\\
&H_{\xi, \text{\,n-l}}\,=\,\frac{1}{8}\,v^2 m_\text{H}^2\,\int
\,(\xi-1)^3(\xi+3)\,d^3r~,
\label{Hnonlin}
\end{align}
then we find, see Fig. \ref{dva}, that the nonlinear part is negative and quite small. Its sign complies with the inequality  $\xi(r) < 1$ observed in Fig. \ref{odin}. Its smallness indicates that an essential contribution to the integral in Eq.(\ref{Hnonlin}) comes from the area where $1-\xi(r)\ll 1$, i. e. the outer surface of the bag, see Fig. \ref{odin}. 
\noindent
\begin{figure}[b]
\centering
\includegraphics[height=2 cm,keepaspectratio = true, 
]{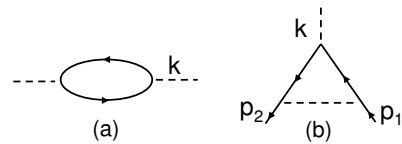}
\caption{
 \label{tri}  Solid and dashed lines - fermion and Higgs fields.}\end{figure}
\noindent

Consider now the radiative corrections for the fermion bag. In the one-loop approximation the most important diagrams are shown in Fig. \ref{tri}. Consider the diagram (a), which represents the variation of the Higgs field due to vacuum polarization produced by fermions. 
Take into account that the deviation of the Higgs wave function $\xi$ from its vacuum expectation value is small. This is evident for small fermion masses, see Fig. \ref{odin} where $1-\xi(r)\ll1$ for $m=350$ GeV. For larger $m$, see the case of $m=600$ in Fig. \ref{odin},  $\xi(r)$ can deviate significantly from 1, but this deviation takes place only in the area of small $r$, which is suppressed in all the integrals over $d^3r$. The smallness of the term $H_{\xi, \text{\,n-l}}$, see Fig. \ref{dva}, is a good example of such suppression. Hence we can assume that the bag does not affect substantially the fermion propagator, which we approximate by its vacuum value. Similarly, the vacuum value is assigned to the Higgs propagator.
Observe also that the size of the bag $R$ is larger than the Compton radius of the fermion, $m R\gg1$. For example, Fig. \ref{odin} shows that for $m=0.35$ TeV and $m=0.6$ TeV the sizes of two bags are roughly $R\approx10$ and $R\approx 5$ TeV$^{-1}$, which yields $mR \approx 3.5$ and $mR\approx 3$ respectively. Therefore, typical transferred momenta within the bag are small, $|\boldsymbol{k}|\sim 1/R \ll m$. We also can assume that $m_\text{H}\ll m$. Consequently, considering the contribution of this diagram, i. e.  the polarization operator $P(k^2)$, we can expand it in powers of $k^2/m^2$ and $m_\text{H}^2/m^2$. According to Ref. \cite{diagrams} the result reads
\begin{equation}
P(k^2)\approx\,-  \nu\,\frac{\,g^2\,(k^2-m_\text{H}^2)^2 }{ 80\pi^2m^2 }=
-\frac{\nu}{8\pi^2}\,\frac{(k^2-m_\text{H}^2)^2}{10 v^2},
\label{P}
\end{equation}
where $\nu$ is the number of heavy fermions with $m \gg m_\text{H}$ (recall the factor $N_c=3$ for quarks).
Remarkably, the large fermion mass, which is present in the coupling constant $g=m/v$ is canceled out in the final expression here and produces no enhancement for $P(k^2)$.
Hence the contributions of all heavy fermions is summed up. For one heavy lepton $\nu=1$, for 
the whole 4-th generation $\nu=8$, and $\nu=11$ when the top quark is also counted (neglecting the contributions of W$^\pm$,  Z$^0$ and Higgs in the loop since condition $m\gg m_\text{H}$ is not valid for them).

Consider now diagram (b) in Fig. \ref{dva}, which describes correction to the vertex 
$\Gamma$ of the fermion-Higgs interaction. 
Take into account that for small fermion masses the binding energy of the fermion is smaller than its mass, $|\epsilon-m|\ll m$. 
From $\epsilon \approx m$ and $|\boldsymbol{k}|\ll m$ we see that the fermion legs of the vertex can be taken on the mass shell, $p_1^2\approx p_2^2\approx m^2$, which makes $\Gamma$ a function of $k^2$ only, $\Gamma=\Gamma(k^2)$. 
The renormalization condition for it reads $\Gamma(0)=g$, which
admits that the radiative corrections are absent for a constant value of the Higgs.
As a result the expansion of $\Gamma(k)$ in powers of $k^2/m^2$ reads
\begin{equation}
\frac{1}{g}~\Gamma(k^2)\,\approx \,1+\frac{g^2}{8\pi^2}\,\gamma\,\frac{k^2}{m^2}
\,=\,1+\frac{\gamma}{8\pi^2} \,\frac{k^2}{v^2}~,
\label{fT}
\end{equation}
where $\gamma$ is an expansion coefficient, which was calculated in \cite{diagrams}. 
Observe that again, see  Eq.(\ref{P}), the large fermion mass
is canceled out in the final expression here and hence does not produce enhancement.
More precisely, $\Gamma(k)/g$ lacks strong, power-type factors of $m$, though its  weaker  dependence on $m$ is present through the coefficient $\gamma$. 
The analytical expression found in \cite{diagrams} for $\gamma$ is lengthy, but it was noted there that a simple fitting
\begin{equation}
\gamma\approx \gamma_\text{\tiny{fit}}={1}/{3}\,\big(\,\ln (1+{1}/{\mu} \,)-
{7}/({4+5\mu})\,\big)~,
\label{fit}
\end{equation}
where $\mu=m_\text{H}/m$, reproduces $\gamma$ with per cent accuracy.

Take next Eq.(\ref{HFG}) and consider there the variation of the field $\delta \xi$ produced by the vacuum polarization due to diagram (a) in Fig. \ref{tri}, as well as the variation of the vertex $\delta g$, see diagram (b). 
The later reads $\delta g= \Gamma(k^2)-g$, while
for the former is
$\delta \xi=(k^2-m_\text{H}^2)^{-1}P(k^2)(\xi-1)$ where we take into account that the polarization changes the deviation of the Higgs field from its vacuum expectation value $\xi-1$. 
Combining the two terms we find the radiative correction to the energy per fermion $\delta\epsilon_\text{rc} \,=\,\langle \bar \psi\,|~\,(1/2)\,g\,v \, \delta \xi  \,+\, \delta\,\Gamma v\, \xi\,|\psi\rangle$. Here the coefficient $1/2$ reflects the fact that two fermions participate in the Higgs boson exchange. Using Eqs. (\ref{P}), (\ref{fT}) we  derive 
\begin{equation}
\delta\epsilon_\text{rc} = \frac{m}{ 8\pi^2 v^2} \langle \,  \bar\psi \,|\,\gamma\,\Delta \xi
-\frac{\nu}{20}\,\big(\Delta\, \xi-m_\text{H}^2\,(\xi-1)\,\big)\,|\,\psi\,\rangle.
\label{delta}
\end{equation}
where we identified $k^2 = \Delta$. 
The terms with the coefficients $\gamma$  and $\nu/20$  
represent the vertex  correction (\ref{fT}) and  vacuum polarization (\ref{P}). 
\begin{figure}[t]
\centering
\includegraphics[height=4.5 cm,keepaspectratio = true, 
]{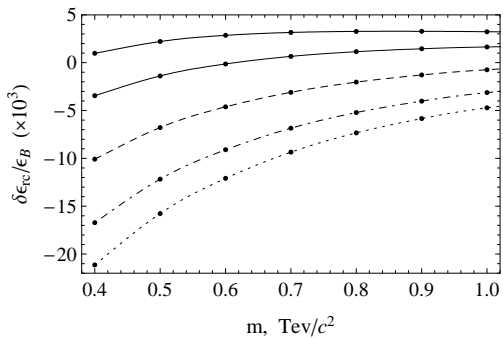}
\caption{
 \label{4etyre} The radiative corrections  (\ref{delta}) scaled by the binding energy $\epsilon_\text{B}$ vs  the fermion mass $m$ for different numbers $\nu$ of heavy fermions  (including quark colors); from top to bottom $\nu=1,3,6,9$ and $11$; the bag is same as in Figs.\ref{odin},\ref{dva}.
 }\end{figure}
\noindent

From Eqs. (\ref{xi}), (\ref{F}), (\ref{fit}), and (\ref{delta}) we find $\delta\epsilon_\text{rc}$. The results plotted in Fig. \ref{4etyre} are scaled by the binding energy $\epsilon_\text{B}=m-\epsilon$, $\epsilon=E/N$, see Fig. \ref{dva}. 
For a hypothetical case $\nu=1$ the dominant contribution to
$\epsilon_\text{rc}$ 
is given by
 the vertex correction, which makes the energy shift positive. For more sensible case of larger $\nu$, in particular $\nu=11$ (complete forth generation plus the top quark), the polarization dominates making the correction negative. However, when $m$ increases the vertex induced contribution rises since $\gamma$ is rising, see (\ref{fit}), which reduces the absolute value of the correction.
The correction is small, $10^{-2}-10^{-3}$, for all $\nu$ and $m$ considered, and is negative for $\nu\ge 6$.

To explain and generalize this result for different numbers of fermions in the bag, $N\le 12$,
we neglect in (\ref{delta}) the small term $\propto m_\text{H}^2 $, and 
make an estimate $\Delta \xi\sim (1-\xi)/R^2$ where $R$ is the bag radius, which is valid since $1-\xi(r)$ varies smoothly. Hence we 
reduce (\ref{delta}) to
\begin{equation}
\frac{\delta \epsilon_\text{rc}}{ \epsilon_\text{B} }\,\approx\,\Big(\gamma-\frac{\nu}{20}\Big)
\frac{1}{4\pi^2 v^2 R^2}\,\approx\,\Big(\gamma-\frac{\nu}{20}\Big)\,\frac{1}{2\pi N}~.
\label{delta/epsilon}
\end{equation}
Here the binding energy is estimated as $\epsilon_\text{B} \approx \frac{1}{2}\,m\,\,\langle \bar\psi\,|\,1-\xi(r)\,|\,\psi \rangle $ (the coefficient $1/2$ comes from the virial theorem). With an increase of $m$ the bag may only shrink, see Fig.\ref{odin}, but
it was shown in \cite{Kuchiev:2008gt} that for sufficiently large $m$ its radius reaches the $m$-independent value $R_0=(N/2\pi)^{1/2}v^{-1}\approx 1.6\,\sqrt N$ TeV$^{-1}$, which was used in the last estimate in (\ref{delta/epsilon}) for $R$.
One verifies numerically that for $N=12$ the simple estimate (\ref{delta/epsilon}) correctly reproduces 
the order of magnitude of the accurate result (\ref{delta}) for the whole range of $\nu$ and $m$ presented in Fig. \ref{dva}.
This estimate shows that the radiative corrections are suppressed as $1/N$ by the number of fermions. An interesting new feature is the small factor $(\gamma-\nu/20)/2\pi\sim 10^{-2}-10^{-1}$, which suppresses the correction, 
$\delta \epsilon/\epsilon_\text{B}\ll 1 $,  
even if only a few fermions $N=2,3\dots$ form the bag.

We omitted a number of processes expecting them to be small. 
The self-energy corrections to the fermions legs are small since the fermions are close to the mass shell. 
The processes with a large number of the Higgs legs are suppressed by small factors $1-\xi$ attached to each leg; compare suppression of the nonlinear processes in Fig. \ref{dva}. 

The problem of radiative corrections in fermion bags show similarities with the atomic Lamb shift. In both cases
the vertex correction gives the repulsive contribution, while 
the vacuum polarization brings in attraction,
in both of them the matrix element of $\Delta U$, $U$ is the nonrelativistic potential energy, plays an important role (in fermion bags $U = m(\xi-1)$). 
For $m\gg m_\text{H}$ the asymptotic $3\gamma \approx \ln (m/m_\text{H})$ from (\ref{fit}) matches Bethe's logarithm $\ln( m_e/B_e)$, $B_e$ is the electron binding energy. Even numerical coefficients in (\ref{delta}) and in the Lamb shift problem are similar. The distinction is that in the bags the number of fermions, which contribute to the vacuum polarization can be large ($\nu=11$), which 
enhances the vacuum polarization making the attraction stronger.

The fermion bags bound by the Higgs boson exchange can exist only if fermions are sufficiently heavy. However, it has been presumed previously that the strong fermion-Higgs coupling makes the radiative corrections so large that they may destabilize the bag. The present work removes this obstacle for the bag formation. We found that the corrections are definitely small for the fermion bags proposed in \cite{Kuchiev:2008gt,Kuchiev:2010ux} provided the fermion mass remains in the region of several hundreds GeV. An encouraging implication is that if there exist fermions of the fourth generation with masses within this interval they certainly are able to form the fermion bags.

Interestingly, the found corrections show no signs of growing with the fermion mass $m$.  One can anticipate therefore that they remain small for masses well above 1 TeV. If correct, this assessment would prove beyond doubt that the fermion bags remain relatively light for very heavy fermions, as was proposed 
in \cite{Kuchiev:2008gt} on the basis of the mean field approach, which gave an estimate $m_\text{bag}\approx 1.2 \sqrt N$ TeV. The bag in this case can be lighter than a fermion, $m_\text{bag}<m$, and hence can be used as a tool for detection of heavy fermions. To justify this physical picture one needs to find a reliable estimate for the radiative corrections for $m>1$ TeV. This problem lies ahead of us. 

Note that different heavy fermions contribute approximately equally 
to the polarization operator and vertex in Eqs.(\ref{P}), (\ref{fT}). This unusual property may be used to measure the number of heavy particles via the Higgs-dependent radiative corrections.


This work is supported by the Australian Research Council.

\end{document}